\documentclass[nonatbib]{sigplanconf}

\usepackage[numbers]{natbib}

\usepackage{float}
\usepackage{eurosym}
\usepackage{enumerate}
\usepackage{graphicx}
\usepackage{multicol}
\usepackage{amsmath}
\usepackage{amssymb}
\usepackage{graphicx}
\usepackage{lipsum}
\usepackage{floatflt}
\usepackage{epsfig}
\usepackage{alltt}
\usepackage{fixltx2e}
\usepackage[table]{xcolor}
\usepackage{subfig}
\usepackage{times}
\usepackage{algcompatible}
\usepackage{algpseudocode}
\usepackage{algorithm}
\usepackage{hhline}
\usepackage{amsmath}
\usepackage{lipsum}
\usepackage{epstopdf}
\usepackage{graphicx}
\usepackage{algpseudocode}
\usepackage{algorithm}
\usepackage{enumerate}
\usepackage{isomath}

\usepackage{amsmath}

\newcommand{\ignore}[1]{}

\newcommand{\boxtheorem}{\hfill $\Box$}
\newcommand{\nit}[1]{{\it #1}}

\def\Li{\mathrm{\Phi}}
\ignore{

}

\newcommand{\comlb}[1]{{\vspace{2mm}\noindent \red{\bf COMM(LEO):}}~ #1 \hfill {\bf    END.}\\}
\newcommand{\red}[1]{\textcolor{red}{#1}}

\ignore{\newcommand{\comlb}[1]{{\vspace{4mm}\noindent \bf COMM(LEO):}~
{\em #1}\hfill {\bf END.}\\}  }

\ignore{
\newcounter{lemmaA-counter}
\newcounter{propositionA-counter}
\setcounter{lemmaA-counter}{0}

\setcounter{propositionA-counter}{0}

{\vskip \abovedisplayskip \refstepcounter{lemmaA-counter}%
\noindent {\bf Lemma A.\arabic{lemmaA-counter}.}}%

{\vskip \abovedisplayskip \refstepcounter{propositionA-counter}%
\noindent {\bf Proposition A.\arabic{propositionA-counter}.}}%
}

\makeatother
\newcommand{\mc}[1]{\mathcal{ #1}}

\newcommand{\blue}[1]{\textcolor{blue}{#1}}
\ignore{
\hypersetup{
  colorlinks   = true, 
  urlcolor     = blue, 
  linkcolor    = blue, 
  citecolor   = red 
}
}

\newcounter{theorem-counter}
\newcounter{corollary-counter}
\newcounter{lemma-counter}
\newcounter{definition-counter}
\newcounter{example-counter}
\newcounter{proposition-counter}
\newcounter{remark-counter}
\newcounter{definitionA-counter}
\newcounter{lemmaA-counter}
\newcounter{propositionA-counter}
{\vskip \abovedisplayskip \refstepcounter{theorem-counter}%
\noindent {\bf Theorem \arabic{theorem-counter}.}}%

\newenvironment{corollary}%
{\vskip \abovedisplayskip \refstepcounter{corollary-counter}%
\noindent {\bf Corollary \arabic{corollary-counter}.}}%

{\vskip \abovedisplayskip \refstepcounter{lemma-counter}%
\noindent {\bf Lemma \arabic{lemma-counter}.}}%

\newenvironment{definition}%
{\vskip \abovedisplayskip \refstepcounter{definition-counter}%
\noindent {\bf Definition \arabic{definition-counter}.}}%

\newenvironment{example}%
{\vskip \abovedisplayskip \refstepcounter{example-counter}%
\noindent {\bf Example \arabic{example-counter}.}}%

\newenvironment{proposition}%
{\vskip \abovedisplayskip \refstepcounter{proposition-counter}%
\noindent {\bf Proposition \arabic{proposition-counter}.}}%

{\vskip \abovedisplayskip \refstepcounter{remark-counter}%
\noindent {\bf Remark \arabic{remark-counter}.}}%

\newcommand{\proof}[1]{{\noindent\bf Proof:\
}#1 \boxtheorem\\ }

\begin{document}
\bibliographystyle{plainnat}

\setlength{\pdfpageheight}{\paperheight}
\setlength{\pdfpagewidth}{\paperwidth}

\ignore{
\conferenceinfo{CONF 'yy}{Month d--d, 20yy, City, ST, Country}
\copyrightyear{2016}
\copyrightdata{978-1-nnnn-nnnn-n/yy/mm}
\copyrightdoi{nnnnnnn.nnnnnnn}

\titlebanner{banner above paper title}        
\preprintfooter{short description of paper}   
}

\title{Quantifying Causal Effects on Query Answering in Databases}

\authorinfo{Babak Salimi}
           {University of Washington, Seattle, USA}
           {bsalimi@cs.washington.edu}
\authorinfo{Leopoldo Bertossi}
           {Carleton University, Ottawa, Canada}
           {bertossi@scs.carleton.ca}

\authorinfo{Dan Suciu}
           {University of Washington, Seattle, USA}
           {suciu@cs.washington.edu}

\authorinfo{Guy Van den Broeck}
           {University of California, Los Angeles, USA}
           {guyvdb@cs.ucla.edu\vspace{-5mm}}

\ignore{\author{{\bf Babak Salimi}, \ {\bf Leopoldo Bertossi}, \ {\bf Dan Suciu} \ and \ {\bf Guy Van den Broeck}}  }
\ignore{ Carleton University, \
School of Computer Science,  Ottawa,  Canada\\}

\maketitle

\vspace{-5mm}
\begin{abstract}
The notion of actual causation, as formalized by Halpern and Pearl, has been recently applied to relational databases, to characterize and compute
{\em actual causes} for possibly unexpected answers to monotone queries. Causes take the form of database tuples, and can be ranked
according to their {\em causal responsibility}, a numerical measure of their relevance as a cause for the query answer. In this work
we revisit this notion, introducing and making a case for an alternative measure of causal contribution, that of {\em causal effect}. In doing so, we  generalize the notion of actual cause, in particular,
going beyond monotone queries. We show that
causal effect provides intuitive and intended results. 
\end{abstract}



\section{Introduction} \label{sec:intro}


The central aim of many scientific disciplines, ranging from philosophy through law and physiology to computer science, is the elucidation of cause-effect relationships among variables or events. In data management in particular, there is a need to represent, characterize
 and compute  causes that explain why certain query results are obtained or not\ignore{, or why natural semantic conditions, such
 as integrity constraints, are not satisfied. Causality can also
 be used to explain the contents of a view, i.e. of a predicate with virtual
 contents that is defined in terms of other physical, materialized relations (tables)}. The notion of causality-based explanation for a query result   was introduced in \cite{Meliou2010a}, on the basis of the deeper concepts of {\em counterfactual} and {\em actual causation}
introduced by Halpern and Pearl in
\citep{Halpern05}, which we call HP-causality. We will refer to this notion
as {\em query-answer causality}, or simply, {\em QA-causality}.

Intuitively, a database atom (or simply, a tuple) \ $\tau$ \ is an {\em actual cause} for an answer $\bar{a}$ to a
monotone query $\mc{Q}$ from  a relational database instance $D$ if there is a ``contingent" subset
of tuples $\Gamma$, accompanying $\tau$,
such that after removing $\Gamma$ from $D$: (a) $\bar{a}$ is still an answer to the query, and (b) further removing $\tau$ from $D\smallsetminus \Gamma$,  makes
 $\bar{a}$ not an answer to the query anymore. (I.e. $\tau$ is a {\em counterfactual cause} under $D\smallsetminus \Gamma$.) \ignore{In other words,
$\tau$ is an actual cause for $\bar{a}$ if there is an intervention (in the form of tuple deletions)
 that makes $\tau$ pivotal for $\bar{a}$. Here ``pivotal" means the query answer counterfactually depends on $\tau$.}

In \citep{Meliou2010a},  the notion of causal responsibility in databases was introduced, to provide a
 metric to quantify the causal contribution, as  a numerical degree, of a
 tuple to a query answer. This responsibility-based ranking is considered as one of the
 most important contributions of HP-causality and its extension \cite{Chockler04} to data management \cite{Meliou2011a}. In  informal terms, causal responsibility as in
 \cite{Chockler04} tells us that, for variables $A$ and $B$,  the degree of responsibility of $A$ for $B$ should be  $\frac{1}{(N + 1)}$,  where $N$ is the {\em minimum number of changes} that have to be made
 on other variables to obtain a situation where $B$
counterfactually, directly depends on $A$. In the case of databases, the responsibility of a cause $\tau$ for an answer $\bar{a}$, is defined as $\frac{1}{1 + |\Gamma|}$, where $\Gamma$ is a smallest-size
contingency set for $\tau$.

Apart from the explicit use of causality, most of the related research on explanations for query
results has concentrated on {\em data provenance} \cite{BunemanKT01,BunemanT07,CuiWW00,Tannen10}\ignore{Cheney09,CuiWW00,Tannen10,tannen}. Causality has been discussed in relation to {\em data provenance} \cite{Meliou2010a,Meliou2011a}
and
{\em workflow provenance} \cite{Cheney10}. Specifically,  in \cite{Meliou2010a}  a close connection between
QA-causality and {\em why-provenance} (in the sense of \cite{BunemanKT01}) was established.

In a different direction, correspondences between causal responsibility
and other concepts and problems in databases, e.g. the {\em view-update problem} and
\ignore{cardinality-based} {\em database repairs} have been established in \cite{Salimi15,uai15,Cibele15,FLAIRS16}. The underlying reason for these connections is
the need to perform a minimal set or minimum number of changes on the database, so that the resulting
state of the database has a desired property. Accordingly, we can see that
actual causality and  causal responsibility are indeed important concepts that may unify several problems in data management.


The notion of causal responsibility as introduced in \cite{Chockler04} has been subject to some
criticism lately \cite{Zultan12,Braham09,Halpern15b,Salimithesis}. \ignore{In \cite{Halpern15b}, it is argued that,
while causal responsibility could capture some reasonable intuitions, alternative definitions might be
more appropriate in some applications.}
In the context of databases,  it has been shown in \cite{Salimithesis}  that causal responsibility
only partially fulfils the original intention of adequately ranking tuples according to their causal contribution
to an answer. We illustrate some of these issues by means of an example  (for others and a discussion, see \cite{Salimithesis}).

\vspace{4mm}
{\small \begin{tabular}{l|c|c|} \hline
$E$~  & ~$A$~ & ~$B$~ \\\hline
$t_1$ & $a$ & $b$\\
$t_2$& $a$ & $c$\\
$t_3$& $c$ & $b$\\
$t_4$& $a$ & $d$\\
$t_5$& $d$ & $e$\\
$t_6$& $e$ & $b$\\ \cline{2-3}
\end{tabular}  }

\vspace{-2.9cm}
\begin{figure}[h]
\hspace*{3.5cm}
\epsfig{file=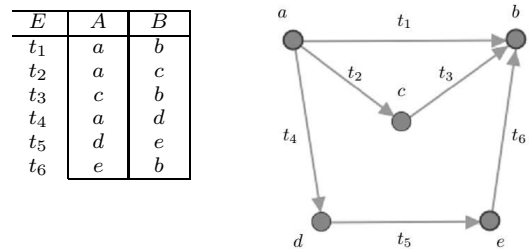,  width=1.6in}

\vspace{-3mm}
  \caption{Instance $D$ and its associated graph $G$ \ignore{in Example \ref{exa:newcr}}}\label{fig:newdc}
\end{figure}

\begin{example} \label{exa:newcrIntro}
Consider instance $D$ with a single binary relation $E$ as in Figure \ref{fig:newdc}. For simplicity, we  use the identifiers, $t_1$-$t_6$, to refer to the database tuples.
Instance $D$ can be represented as the directed graph $G(\mc{V}, \mc{E})$ in Figure \ref{fig:newdc}, where $\mc{V}$ is the active domain of $D$, and $\mc{E}$ contains an edge $(v_1, v_2)$ iff $E(v_1, v_2) \in D$.  Tuple identifiers are used as labels for the corresponding edges in the graph.

Now consider query, $\mc{Q}$  asking if there exists a path between $a$ and $b$. This query is monotone, Boolean (i.e. it has a \nit{true}/\nit{false} answer), and can be expressed
in recursive Datalog.  The answer is \nit{true} in $D$. All the tuples are actual causes for this answer, with the same causal responsibility: \ $\frac{1}{3}$. \
However, since $t_1$ provides a direct connection, it makes sense to claim that $t_1$
contributes more to this answer than the other tuples. Intuitively, tuples that belong to shorter paths between $a$
and $b$ contribute more to the answer than tuples that belong to longer paths.
\boxtheorem
\end{example}

\vspace{1mm}
In this work we introduce the notion of
causal effect in the context QA-causality in databases. Causal effect refers to the extent to which an input variable has a direct influence or drive on the next state
of a output variable, i.e. addressing questions of the form: \ ``If we change the state of the
input, to what extent does that alter the state of the
output?".
To achieve this goal, we start from the central notion of {\em causal effect} that is used in the
{\em theory of causal inference} proposed in \cite{PearlBook}. Introducing it in the context QA-causality in databases allows us to define and investigate
the notion of {\em causal effect of a tuple on the answer to a  Boolean query}.

\ignore{
Following the general approach to causal effect in causal inference, we define causal effect in our QA-setting by constructing {\em structural causal models} \cite{PearlBook,Pearl09},
which in our case become essentially propositional models involving random variables associated to  (the presence or absence of) database tuples (in/from the database at hand).
These models are obtained from the {\em lineages} of the queries that may involve negation \cite{Dan11}.}

We show that, in databases, causal effect subsumes
the notion of actual causal as introduced in \cite{Meliou2010a}, and can be sensibly applied not only to
monotone queries, but also to first-order queries with negation,  and to aggregate queries. Furthermore, we illustrate, by means of several examples, that this notion provides an intuitive and informative alternative to causal responsibility when ranking causes according to their causal contribution to a query result.

\ignore{
Actual
causation is a binary notion, i.e. for a given query and instance, a tuple is either an actual cause or not. However,  causal effect is
inherently a numerical, graded measure, and immediately ranks tuples in terms of their causal contribution
to a query answer. For an example, as we will see later on,  causal effect, applied to the instance and query in Example \ref{exa:newcrIntro}, provides
 the following causal effect values for
  the tuples $t_1, \ldots, t_6$: \ $0.65625$, \ $0.21875$, \ $0.21875$, \ $0.09375$, \ $0.09375$, \ $0.09375$, respectively. This result perfectly matches our intuition.
}
\ignore{
 We show in technical terms that for Boolean queries and linear aggregate queries, the notion of causal effect is related to the Pearson correlation coefficient.
 as a natural extension of the notion of actual causation for query answers. As oppose to the notion of actual causation
}

\ignore{In Section \ref{sec:cdepend}, we extend the notion of QA-causality by considering
notions  of {\em sufficiency} and {\em necessity} that will be used later in the chapter to define a {\em degree of causal contribution}. In addition,  we provide a couple of examples where causal responsibility
returns possibly non-intuitive results. Section \ref{sec:discDegree} propose a new metric for numerical quantification of causal contribution, and investigate its properties.
}

\vspace{-0.2cm}
\section{Preliminaries} \label{sec:pre}



We consider relational database schemas of the form $\mathcal{S} = (U,  \mc{P})$,   where $U$ is a finite
database domain of {\em constants} and $\mc{P}$ is a finite set of {\em database predicates} of fixed arities. In some cases, we may also have {\em built-in} predicates, e.g. $\neq$, that we
leave implicit.  A database instance $D$
compatible with $\mathcal{S}$ can be seen as a finite set of ground atomic formulas   of the form $P(c_1,   ...,   c_n)$,   where $P \in \mc{P}$ has arity $n$,   and $c_1,   \ldots ,   c_n \in U$.  In databases
 these formulas are usually called atoms or {\em tuples}. They will be denoted with $\tau, \tau_1, \ldots$. The active domain of an instance $D$, denoted $\nit{Adom}(D)$, is the finite set of constants from $U$ that appear in $D$. \ignore{\blue{In the following, we will assume that, for every instance $D$,
$\nit{Adom}(D) \subseteq \nit{Dom}$, where the latter is fixed and finite subset of  $U$.}}

In this work, we will mostly consider first-order (FO) queries, that is, formulas $\mc{Q}(\bar{x})$ of the language of FO predicate logic,   $\mc{L}(\mc{S})$,   associated to $\mc{S}$. In $\mc{Q}(\bar{x})$,  \ $\bar{x}$    shows
all the free variables in the formula. If $\bar{x}$ is non-empty,   the query is {\em open}. If $\bar{x}$ is empty,   the query is {\em Boolean},   i.e. a sentence,   in which case,   the answer
is \nit{true} or \nit{false}
in a database,   denoted by $D \models \mc{Q}$ and $D \not\models \mc{Q}$,   respectively. A sequence $\bar{c}$ of constants is an answer to an open query $\mc{Q}(\bar{x})$ if $D \models \mc{Q}[\bar{c}]$,   i.e.
the query becomes true in $D$ when the variables are replaced by the corresponding constants in $\bar{c}$. We denote with $\mc{Q}(D)$ the set of all answers to query $\mc{Q}(\bar{x})$.

 In particular, a {\em conjunctive query} \ (CQ) is a formula  of the form \ $\mc{Q}(\bar{x})\!: \ \exists \bar{y}(P_1(\bar{s}_1) \wedge \cdots \wedge P_m(\bar{s}_m))$,
where the $P_i(\bar{s}_i)$ are atomic formulas,   i.e. $P_i \in \mc{P}$,   and the $\bar{s}_i$ are sequences of terms,   i.e. variables or constants.\footnote{We say it explicitly when we allow the
 $P_i$ to be built-ins.} When $\bar{x}$ is empty,   the query is {\em Boolean} conjunctive query (BCQ).

A query $\mc{Q}$ is {\em monotone} if for every two instances $D_1 \subseteq D_2$, \ $\mc{Q}(D_1) \subseteq \mc{Q}(D_2)$, i.e. the set of answers grows monotonically with the
instance. For example,  {CQ}s and unions of  {CQs} ({UCQ}s) are monotone. Datalog
queries \cite{Abiteboul95}, although not always expressible as FO queries, are also monotone. Although most of the work on QA-causality has concentrated on monotone queries, in this work
we will also consider non-monotone queries.

Now we review the notions of QA-causality and responsibility as introduced in \cite{Meliou2010a}.
Assume the relational instance $D$ is split in two disjoint sets: $D=D^n \cup D^x$, where $D^n$ and $D^x$ are the sets of {\em endogenous} and {\em exogenous} tuples,
respectively.\footnote{ Endogenous tuples
are admissible, possible candidates for causes, as opposed
to exogenous tuples. The  partition is application-dependent and captures predetermined factors, such as users' preferences that may affect QA-causal analysis.} Let $\mc{Q}$ be a monotone Boolean query. A tuple $\tau \in D^n$ is  a
{\em counterfactual cause} for an answer $\mc{Q}$ in $D$  if $D\models \mc{Q}$,
but $D\smallsetminus \{\tau\}  \not \models \mc{Q}$.  A tuple $\tau \in D^n$ is an {\em actual cause} for  $\mc{Q}$
if there  exists $\Gamma \subseteq D^n$, called a {\em contingency set}, such that $\tau$ is a
counterfactual cause for $\mc{Q}$ in $D\smallsetminus \Gamma$. The {\em causal responsibility} of a tuple $\tau$ for answer $\bar{a}$, denoted  $\rho_{_{\!\mc{Q}\!}}(\tau)$, is $\frac{1}{1 + |\Gamma|}$, where $|\Gamma|$ is a
smallest-size contingency set for $\tau$. When $\tau$ is not an actual cause for $\bar{a}$,
no contingency set is associated to $\tau$. In this case, $\rho_{_{\!\mc{Q}\!}}(\tau)$ is defined as $0$. \ignore{
Notice that causal responsibility subsumes
actual causation, in the sense that: \ $\tau$ is an actual cause for $\mc{Q}$ iff $\rho_{_{\!\mc{Q}\!}}(\tau)>0$.}
\ignore{
Notice that the definitions
of actual cause,  and responsibility apply,  in particular, to
 {\em unions of BCQs} (UBCQs), with or without built-ins, and Datalog queries, possibly with recursion \cite{uai15}.}


\subsection{Lineage of a query} \label{sec:lineage}

\ignore{
Lineage of  queries will be used in Section \ref{sec:degree} as the basis for a structural causal models. Those models will allow us to express logical relationships between variables that express the
presence or absence of associated  tuples in or from a database instance at hand. We discuss in more detail the case of first-order queries.}

The {\em lineage (expression)} of a Boolean FO query $\mc{Q}$, as used in probabilistic databases \cite{Dan11},  is a propositional formula, $\Li_\mc{Q}$, over the finitely many potential tuples in
an arbitrary database instance for the schema at hand, i.e. all tuples \ $\tau\!: P(c_1,\ldots,c_n)$, with $n$-ary $P \in \mc{P}$ and $c_1, \ldots, c_n \in U$. For each such a $\tau$, we introduce a
propositional a propositional variable $X_\tau$ (aka. a propositional atom). $\nit{Var}(\mc{S}, U)$ denotes the set of variables associated to tuples. It depends on the
schema and data domain, and determines a propositional language $\mc{L}(\nit{Var}(\mc{S}, U))$.

Formula $\Li_\mc{Q}$ expresses which input tuples must be present in the database and which tuples must be absent from it for the query to be true. \ignore{
The lineage of  $\mc{Q}$ on $D$ is the propositional formula $\Li^D_\mc{Q}$,
or simply} $\Li_\mc{Q}$ \ignore{if $D$ is understood from the context,} is defined inductively for first-order (FO) queries $\mc{Q}$, as follows:
 \ 1. \ If $\mc{Q}$ is a
tuple $\tau$, $\Li_\tau := X_\tau$.
\ 2. \ $\Li_{a=a}:=\nit{\tiny true}$. \ \ 3. \ $\Li_{a=b}:=\nit{\tiny false}$. \ \ 4. \ $\Li_{\mc{Q} \land \mc{Q}_2}:=\Li_{\mc{Q}_1} \land \Li_{\mc{Q}_2}$.  \ 5. \ $\Li_{\mc{Q}_1 \lor \mc{Q}_2}:=\Li_{\mc{Q}_1} \lor \Li_{\mc{Q}_2}$. \
  \ 6. \ $\Li_{\exists x \mc{Q}}:= \bigvee\limits_{c \in U} \Li_{\mc{Q}[\frac{c}{x}]}$. \ \ 7. \ $\Li_{\neg \mc{Q}}:= \neg(\Li_\mc{Q})$.

\ignore{In 5., \ $\mc{Q}[\frac{c}{x}]$ denotes the query resulting from replacing variable $x$ by constant $c$ in $\mc{Q}$.} For a query $\mc{Q}$, $\nit{Var}(\Li_\mc{Q})$ denotes the set of propositional variables in $\Li_\mc{Q}$. Clearly,  $\nit{Var}(\Li_\mc{Q}) \subseteq  \nit{Var}(\mc{S}, U)$.

\ignore{Notice that this definition also applies to non-monotonic FO queries that involve negation. 
Also notice that every instance $D$ for the schema defines a truth assignment (or valuation) $\sigma_{\!\!_{D}}$ for $\mc{L}(\nit{Var}(\mc{S}, U))$: \ $\sigma_{\!\!_{D}}(X_\tau) = 1$ \ iff \ $\tau \in D$. It holds: \ $D \models \mc{Q} \ \mbox{ iff } \ \sigma_{\!\!_{D}} \models \Li_\mc{Q}$.}

We can make the lineage of a query depend on the instance $D$ at hand, by assigning -at least conceptually- truth values to some of the variables appearing in $\Li_\mc{Q}$ depending on the contents of $D$. More specifically, under the assumption that negation appears only in front of propositional variables (producing negative {\em literals}), the $D${\em -lineage} of $\mc{Q}$, denoted $\Li_\mc{Q}(D)$ is obtained from $\Li_\mc{Q}$ by: \ (a) Making \nit{false} each positive occurrence of variable $X_\tau$ for which $\tau \notin D$. \ (b)  \ Making each  $\neg X_\tau$ \nit{false} for which $\tau \in D$. \ We denote with $\nit{Var}^D(\Li_\mc{Q}(D))$ the set of variables in $\Li_\mc{Q}(D)$. \ {\em We will assume that: (a) a variable $X_\tau$ never appears both positively and negatively in $\Li_\mc{Q}(D)$; and (b) every $\tau \notin D$ that appears negatively in $\Li_\mc{Q}(D)$ is considered to be endogenous.} (Then, endogenous tuples are those in $D^n$ plus some  outside the instance at hand.)

\ignore{In Example \ref{ex:nonmon}, we obtain the following ``instantiated" lineage: \ \ $\Li_\mc{Q}(D)= (X_{R(a, b)} \wedge \neg X_{S(b)})  \vee (X_{R(c, b)} \wedge \neg X_{S(b)})$,
and $\nit{Var}^D(\Li_\mc{Q}(D)) = \{X_{R(a, b)}, X_{S(b)}, X_{R(c, b)}, X_{S(b)} \}$.}

\vspace{-0.1cm}
\begin{example} \label{ex:nonmon}
Consider an schema with two relations,  $R(A,B)$ and $S(B)$,  instance $D=\{R(a,b),R(a,c),R(c,b),S(c)\}$, with data domain $U = \{a,b,c\}$, and the Boolean query $\mc{Q}: \ \exists x (R(x, y) \wedge \neg S(y))$,
which has answer \nit{true} in $D$.  We obtain the following ``instantiated" lineage: \ \ $\Li_\mc{Q}(D)= (X_{R(a, b)} \wedge \neg X_{S(b)})  \vee (X_{R(c, b)} \wedge \neg X_{S(b)})$, with
$\nit{Var}^D(\Li_\mc{Q}(D)) = \{X_{R(a, b)}, X_{S(b)}, \linebreak X_{R(c, b)}, X_{S(b)} \}$.\boxtheorem \end{example}

\ignore{\ \ Here,
\begin{eqnarray*}
\hspace*{-6mm}\Li_\mc{Q}&=& \Li[X_{R(a, a)} \wedge \neg X_{S(a)}] \vee \Li[X_{R(a, b)} \wedge \neg X_{S(b)}]\\&& \hspace{1cm}\vee  \cdots \vee \ \Li[X_{R(c, c)} \wedge \neg X_{S(c)}]\\
&=& (X_{R(a, a)} \wedge \neg X_{S(a)})  \vee (X_{R(a, b)} \wedge \neg X_{S(b)})\\&& \hspace{1cm}  \vee \cdots \vee \ (X_{R(c, c)} \wedge \neg X_{S(c)}).\\
\nit{Var}(\Li_\mc{Q}) &=& \{X_{R(a, a)}, X_{R(a, b)}, X_{R(a, c)}, X_{R(b, b)}, X_{R(b, a)},\\&& \hspace*{-1.5cm}X_{R(b, c)}, X_{R(c, c)},
X_{R(c, a)}, X_{R(c, b)}, X_{S(a)},  X_{S(b)}, X_{S(c)}\}.
\end{eqnarray*}}
\ignore{ Notice that not all possible tuples appear represented by propositional variables in $\nit{Var}(\Li_\mc{Q})$, e.g. $X_{E(d,a)} \in \nit{Var}(\mc{S}, U) \smallsetminus \nit{Var}(\Li_\mc{Q})$.}
\ignore{
This lineage expression does not depend on the instance. However, the lineage becomes \nit{true} under the assignment $\sigma_{\!\!_{D}}$, corresponding to the fact that
$\mc{Q}$ is true in $D$.}

\vspace{1mm}For monotone FO queries, this instantiated lineage corresponds to the {\sf PosBool} provenance semi-ring \cite{Tannen10}, and is related to the {\em minimal-witness-basis}, or {\em why-provenance} \cite{BunemanKT01}.
Notice that lineage can be naively extended to Datalog queries by considering the ground instantiation of a program and disjunctively collecting paths from the query goal all the way
down, through backward-propagation via the ground propositional rules, to the ground
extensional tuples $\tau$, for which variables $X_\tau$ are introduced.

\section{\ Interventions and Causal Effect}\label{sec:degree}

HP-causality \cite{Halpern05}, which is the basis for the notion of QA-causality in \cite{Meliou2010a}, provides a ``structural" model of actual causation. \ignore{It is a framework to express and discover causal relationships in domains where all aspects (variables) of the world are known, and no uncertainty is involved.\footnote{Causality has also been modeled in probabilistic domains \cite{PearlBook}. However, we assume here that
under the HP-model, there are no {\em intrinsic or explicit} probabilities involved.} }  According to that approach,  a {\em causal model} of a particular domain is represented in terms of variables, say $A, B, ...$, their values, and a set of {\em structural equations} representing causal relationships between
variables \cite{Halpern05}.  In this context, the statement ``$A$ is an actual cause for  $B$" claims that there is a set of possible {\em interventions}
 (or contingencies) on the causal model that makes $B$ {\em counterfactually} depend on $A$. That is, had $A$ not happened, $B$ wouldn't have happened.

\ignore{
Modeling with structural equations has become a prominent approach to causal inference \cite{PearlBook}, where the task is to learn causal relationships from raw data on the basis of certain qualitative causal assumptions
 that are deemed plausible in a given domain.}

In  QA-causality,  counterfactual questions take concrete forms, such as: ``What would be (or how would change) the answer to a  query $\mc{Q}$ if the tuple $\tau$ is deleted/inserted from/into the database $D$?" \ A question like this can be addressed by building a corresponding causal model, which, for a query $\mc{Q}$ and instance $D$, becomes the combination of
the query lineage $\Li_\mc{Q}(D)$ and the truth assignment $\sigma_{\!\!_{D}}$ determined by $D$  ($X_\tau$ is true iff $\tau \in D$). This models captures the causal relationships between database tuples (or their propositional variables)
and $\mc{Q}$.

\ignore{More specifically,
in this causal model, the variables
become the $X_\tau$ appearing in the lineage expression, and the latter serves as the structural equation that reflects the causal relationship between the propositional variables that appear in the lineage (whose values are determined by the instance) and the query answer.}

The interventions that represent counterfactual hypothesis become, in this context, insertions or deletions of database tuples $\tau$, which change the truth values originally assigned by $\sigma_{\!\!_{D}}$
to the propositional variables $X_\tau$ appearing in the query lineage $\Li_\mc{Q}(D)$.\ignore{\footnote{In the case of Datalog queries, interventions are similarly limited to changes of values
on ``extensional" variables only, i.e. those associated to extensional tuples.}}

Informally for the moment, interventions will be assignments (or changes) of truth values to (some of) the variables in the lineage.
At some point later on, we will deal with the truth values assigned to variables in $\nit{Var}^{x}(\Li_\mc{Q}(D))$, the set of variables in $\nit{Var}(\Li_\mc{Q}(D))$
  corresponding to exogenous tuples. Positive ``exogenous variables" in  $\Li_\mc{Q}(D)$ form the set $\nit{Var}^{x,+\!}(\Li_\mc{Q}(D))$, and negative ones, the set $\nit{Var}^{x,-\!}(\Li_\mc{Q}(D))$.  Instead of dealing with these variables at the lineage level, we will consider the values
  interventions assign to them \ (see (\ref{eq:exo}) below).

  Now, an {\em intervention on an instance $D$ wrt. a Boolean query $\mc{Q}$} can be represented by a truth assignment $\sigma\!:  \nit{Var}(\Li_\mc{Q}(D)) \rightarrow \{0,1\}$.  The intervention $\mc{I}_\sigma$ on $D$ associated
  to $\sigma$ is the restriction of $\sigma$ to those variables $X_\tau$ such that $\sigma(X_\tau) \neq \sigma_{\!\!_{D}}(X_\tau)$. That is,  $\mc{I}_\sigma$ represents only the changes of truth values,
  i.e. insertions or deletions of tuples into/from $D$. Then, the set of variables that change values wrt. $D$ becomes the domain of $\mc{I}_\sigma$.

  If we consider that assignments can be randomly and uniformly chosen, we obtain a probability space, with {\em outcome space} $\Omega = \{\sigma~|~\sigma: \nit{Var}(\Li_\mc{Q}(D)) \rightarrow \{0,1\}\}$,
  and $P$ the uniform distribution.

Next, we can use Pearl's notation  for {\em interventions} \cite{PearlBook}, i.e.  expressions of the form $\nit{do}(X_\tau$ $ =$ $ x)$, where $x \in \{0,1\}$. It denotes the intervention that makes $X_\tau$ take value $1$ (i.e. becomes {\it true}) or $0$ (i.e. becomes {\it false}), corresponding to
inserting or deleting $\tau$ into/from a database instance, respectively. This notation can be generalized for multiple, simultaneous interventions, with the obvious meaning, to: \ $\nit{do}(\bar{X}) = \bar{x}$, where $\bar{X} \subseteq \nit{Var}(\Li_\mc{Q}(D))$ is a list of $m$ different variables,
and $\bar{x} \in \{0,1\}^m$.
More technically, an intervention $\nit{do}(\bar{X} = \bar{x})$ becomes an {\em event} (a subset of $\Omega$): \ 
$\nit{do}(\bar{X} = \bar{x}) := \{\sigma \in \Omega~|~ \ignore{\nit{Var}(\sigma) = \bar{X} \mbox{ and } }
(\sigma(X))_{X \in \bar{X}} = \bar{x}\}$. 

Query $\mc{Q}$ can be seen as a Bernouilli {\em random variable}: \ $\mc{Q}\!: \Omega \rightarrow \{0,1\}$, defined by \ $\mc{Q}(\sigma) = 1 \mbox{ iff } \sigma \models \Li_\mc{Q}$.
Accordingly, for $y \in \{0,1\}$, we may consider the event ``$\mc{Q} = y" := \{\sigma \in \Omega~|~ \mc{Q}(\sigma) = y\}$. Furthermore, we obtain properly defined
conditional probabilities of the form $P(\mc{Q}=y~|~\nit{do}(\bar{X} = \bar{x}))$.

For a tuple $\tau$ and a value $v$ for $X_\tau$, we can compute the so-called  {\em interventional conditional expectation} of (the truth value of) $\mc{Q}$, namely: \ $E(\mc{Q}~|~\nit{do}(X_\tau = v))
= P(\mc{Q} = 1|~\nit{do}(X_\tau = v))$.\ignore{\footnote{Which happens to be a regular conditional expectation in our database setting where presence/absence of (extensional) tuples are
the only direct causes for query answers, and where ``\nit{do}" interventions can be represented as regular events.}}

In database causality, some tuples are endogenous and others exogenous, but our assignments $\sigma$ on the set of variables do not make such a distinction.
In the following, the expected value will be conditioned on the exogenous variables (those in $\nit{Var}^x(\Li_\mc{Q}(D))$) taking the value $1$ when positive, and value $0$ when
negative. Accordingly, for
an endogenous tuple $\tau$, we redefine:
\begin{eqnarray}
 E(\mc{Q}~|~\nit{do}(X_\tau = v)) &:=& E(\mc{Q}~|~ \ \ \nit{do}(X_\tau = v) \ \cap \label{eq:exo}\\ && \hspace*{-3.5cm}\bigcap_{X_{\tau'} \in \nit{Var}^{x,+\!}(\Li_\mc{Q}(D))} \!\!\!\!\!\!\!\!\!\!\!\!\! \nit{do}(X_{\tau'} = 1) \ \cap \ \!\!\!\!\!\!\!\!\!\! \bigcap_{X_{\tau'} \in \nit{Var}^{x,-\!}(\Li_\mc{Q}(D))} \!\!\!\!\!\!\!\!\!\! \nit{do}(X_{\tau'} = 0 \ )). \nonumber
 \end{eqnarray}
 In the following, we assume that conditional expectations are conditioned on exogenous tuples as in (\ref{eq:exo}). We can now define a {\em measure of the causal effect of an intervention}  \cite{PearlBook} in terms of the average difference between the effects of
two interventions.

\begin{definition}\label{def:causalEff}  Let $D$ be an  instance, $\mc{Q}$ a Boolean FO query\ignore{ with $D \models \mc{Q}$}, and  $\tau \in D^n$.
  The {\em causal effect} of tuple $\tau$ on  $\mc{Q}$ \ignore{(being true)} in $D$ is: \vspace{-1mm}
\begin{equation}
\mathcal{E}^D_{\tau,\mc{Q}} \ := \ E(\mc{Q}~|~\nit{do}(X_\tau = v)) - E(\mc{Q}~|~\nit{do}(X_\tau = 1-v)), \vspace{-4mm} \label{eq:ce}
\end{equation}
\phantom{po}\\
\noindent where $v = 1$ if $\tau \in D$, and $v=0$ if $\tau \notin D$. \boxtheorem
\end{definition}

\vspace{1mm}Intuitively, $\mathcal{E}^D_{\tau,\mc{Q}}$ shows how deleting an existing tuple from instance $D$ or inserting an absent tuple into $D$  affects the mean of the distribution of $\mc{Q}$ (the expectation taken in the space of random interventions on the remaining tuples).

\begin{proposition} \label{pro:causesuffset}  Let $D$ be an  instance, $\mc{Q}$ a Boolean FO query\ignore{ with $D \models \mc{Q}$}, and  $\tau \in D^n$.
\ It holds \ $\mathcal{E}^D_{\tau,\mc{Q}}\ \geq 0$. \boxtheorem
\end{proposition}

\vspace{1mm}We will say that {\em a tuple has a causal effect on} $\mc{Q}$ in $D$ when  $\mathcal{E}^D_{\tau,\mc{Q}}>0$. Causal effect allows us to compare the causal contribution of
tuples: $\tau$ has {\em higher causal effect} on $\mc{Q}$ than $\tau'$ if $\mathcal{E}^D_{\tau,\mc{Q}} > \mathcal{E}_{\tau',\mc{Q}}$. Notice than the definition of causal effect
does not require the query to be true in the given instance $D$. \ We claim that causal effect captures the notion of actual cause.

\ignore{
\begin{definition} \label{def:causesuffset}   Let $D$ be an  instance, $\mc{Q}$ a Boolean query with $D \models \mc{Q}$, and  $\tau, \tau' \in D^n$.
\begin{itemize}
    \item[(a)] $\tau$ has a  {\em positive causal effect} on $\mc{Q}$ if $\mathcal{E}^D_{\tau,\mc{Q}}>0$; a
     {\em negative causal effect} on $\mc{Q}$ if $\mathcal{E}^D_{\tau,\mc{Q}}<0$; and {\em no causal effect} on $\mc{Q}$ if
    $\mathcal{E}^D_{\tau,\mc{Q}}=0$.
    \item[(b)] $\tau$ has {\em more causal effect} on $\mc{Q}$ than $\tau'$ if $|\mathcal{E}^D_{\tau,\mc{Q}}| \geq |\mathcal{E}^D_{\tau',\mc{Q}}|$. \boxtheorem
\end{itemize}
\end{definition}
\begin{proposition} \label{pro:causesuffset}   Let $D$ be an  instance, $\mc{Q}$ a monotone Boolean query with $D \models \mc{Q}$, and  $\tau \in D^n$.
\ It holds \ $\mathcal{E}^D_{\tau,\mc{Q}}\ \geq 0$. \boxtheorem
\end{proposition}  }


\begin{proposition} \label{pro:causesuffset2}  Let $D$ be an  instance, $\mc{Q}$ a monotone Boolean FO query with $D \models \mc{Q}$, and $\tau \in D^n$. \ It holds: \
$\tau $ is an actual cause for $\mc{Q}$ in $D$ \ iff  \ $\tau$ has positive causal effect on $\mc{Q}$ in $D$. \boxtheorem
\end{proposition}
\begin{example} \label{exa:eff} (ex. \ref{exa:newcrIntro} cont.)
We can compute the  causal effects of tuples in $D$ on the query $\mc{Q}$ asking if there is a path between $a$ and $b$. \
Here,
$\Li_\mc{Q}(D)= X_{t_1} \lor (X_{t_2} \land X_{t_3}) \lor (X_{t_4} \land X_{t_5} \land X_{t_6})$. \ Assuming all tuples are endogenous, $\mathcal{E}^D_{t_1,\mc{Q}}=0.65625$,
$\mathcal{E}^D_{t_2,\mc{Q}}=\mathcal{E}^D_{t_3,\mc{Q}}=0.21875$, and $\mathcal{E}^D_{t_4,\mc{Q}}=\mathcal{E}^D_{t_5,\mc{Q}}=
\mathcal{E}^D_{t_6,\mc{Q}}=0.09375$.
\boxtheorem
\end{example}

\vspace{1mm}The notion of causal effect can handle non-monotone queries.

\begin{example} \label{exa:ex:nonmoneffect} (ex. \ref{ex:nonmon} cont.) \ Here,
$\Li_\mc{Q}(D)= (X_{R(a,b)} \land \neg X_{S(b)}) \lor (X_{R(v,b)} \land \neg X_{S(b)})$. \ If all tuples are endogenous,
\
$\mathcal{E}^D_{R(a,b),\mc{Q}}=\mathcal{E}^D_{R(c,b),\mc{Q}} = 0.25$, and $\mathcal{E}^D_{\neg S(b),\mc{Q}}=0.75$.
\ignore{
We observe that
$S(b)$ has a negative causal effect on $\mc{Q}$, meaning that inserting it into the database has a negative causal impact on $\mc{Q}$.
One may interpret this as the absence of $S(b)$ (or $\neg S(b)$ being true) having a positive causal effect on the query, and its causal effect is
greater
than those of other tuples.}
\boxtheorem
\end{example}

\vspace{1mm}Our next example shows that  causal effect can be applied to queries with aggregation. First notice that, in order
to compute the effect of intervention, we do not have to materialize and process the lineage. Each intervention
specifies an instance to which the query can be posed
and evaluated. This allows us to naturally extend  causal effect to aggregate queries. \ignore{\footnote{Provenance, that is related but normally more complex than lineage, for aggregate queries has been investigated
in \cite{tannen11}.}}

\begin{example} \label{exa:agg1} Consider an instance $D$ with a unary relation $R = \{450, 150, 100, -100\}$,
an the Boolean query

 $\mc{Q}$: \verb|select 'true' from R having sum(A) > 500|, asking if the sum of
values in $R$ is greater than $500$.
This  non-monotone query (tuple insertions may invalidate a previous answer) has
 answer \nit{true}, with all numbers
in $R$  contributing to it, but with different causal effects: \
$\mathcal{E}^D_{R(450),\mc{Q}}=0.625, \mathcal{E}^D_{R(150),\mc{Q}}=0.375, \mathcal{E}^D_{R(100),\mc{Q}}=0.125$, and
$\mathcal{E}^D_{R(-100),\mc{Q}}=-0.125$. The negative effect of $R(-100)$ means the tuple has a negative causal impact on the query outcome,  which is intuitive.

Now consider the query  \ $\mc{Q}'\!:$ \ \verb|select AVG(A) from R|.  \ Here, \
$\mathcal{E}^D_{R(450),\mc{Q}'}=112.5, \ \mathcal{E}^D_{R(150),\mc{Q}'}=37.5, \ \mathcal{E}^D_{R(100),\mc{Q}'}=25$, and \
$\mathcal{E}^D_{R(-100),\mc{Q}'}=-25$. \boxtheorem
\end{example}

\newpage

Finally, we point out that  causal effect can be applied to Datalog queries, as that in Example \ref{exa:newcrIntro}, where we obtain: \ $\mathcal{E}^D_{t_1,\mc{Q}}=0.65625$,
$\mathcal{E}^D_{t_2,\mc{Q}}=\mathcal{E}^D_{t_3,\mc{Q}}=0.21875$, and $\mathcal{E}^D_{t_2,\mc{Q}}=\mathcal{E}^D_{t_2,\mc{Q}}=
\mathcal{E}^D_{t_2,\mc{Q}}=0.09375$.

\section{\ Causal Effect and Pearson Correlation}\label{sec:pearson}

In Statistics, the Pearson's {\em correlation coefficient} is a measure of the linear dependence between two random variables $X$ and $Y$. It is  defined by \ $r_{X,Y}=\frac{\nit{Cov}(X,Y)}{\sigma_X \sigma_Y}$, where
$\nit{Cov}(X,Y) := E((X - \mu_X)(Y - \mu_Y))$ is the covariance of $X, Y$, $\mu_X, \mu_Y$ are the expected values of $X, Y$, and $\sigma_X, \sigma_Y$ their standard deviations.

It turns out that there is a close numerical connection between casual effect as introduced above and Pearson's correlation coefficient. This follows from the fact that the probability of any propositional
formula, so as its conditional probability on a given variable, is a multi-linear polynomial in its variables \cite{Dan11}.

\begin{proposition} \label{pro:corr}   Let $D$ be an  instance, $\mc{Q}$ a Boolean FO query. \ignore{ with $D \models \mc{Q}$, and  $\tau \in D^n$} It holds: \ (a) \  If $\tau$ is endogenous and $X_\tau$ appears positively
in $\Li_\mc{Q}(D)$: \ \ $\mathcal{E}^D_{\tau,\mc{Q}} \ = \ \ r_{\mc{Q},X_\tau} \times \frac{\sigma_\mc{Q}}{\sigma_{X_\tau}}$.  (b) If $\tau$ is endogenous and appears negatively
in $\Li_\mc{Q}(D)$: \ \ $\mathcal{E}^D_{\tau,\mc{Q}} \ = \ - r_{\mc{Q},X_\tau} \times \frac{\sigma_\mc{Q}}{\sigma_{X_\tau}}$. Here,  $\mc{Q}$ and $X_\tau$ are treated as Bernouilli random variables on space $\Omega$. \ignore{ $X_\tau$  takes value
$1$ on an assignment $\sigma \in \Omega$ \ iff \ $\sigma(X_\tau) = 1$).} \boxtheorem
\end{proposition}

\vspace{1mm} Unlike causal effect, Pearson's correlation coefficient is a normalized
real-valued measure. For monotone queries, it takes values  between $0$ and $1$\ignore{ (see Example \ref{exa:agg1})}. Then, we may
use this correlation coefficient as a
 measure of the normalized causal effect of a tuple on a query.

\ignore{
Since the lineage of a Boolean Datalog queries can also be represented as a propositional formula, the definitions and results above apply to that case too.}
\ignore{
It is easy to see that the causal effect of $\tau$ on $\mc{Q}$ is \blue{$a-b$}, the slope of the regression line expressing query $\mc{Q}$ as a random variable in terms of
 random variable $X_\tau$.\\
 Query $\mc{Q}'$ in Example
 \ref{exa:agg1} clearly illustrates this, because one can write:
 {\small $$\mc{Q}'= \frac{450 \times X_{450}  +  150 \times X_{150}  +  100 \times X_{100}  +  (-100) \times X_{-100}}{4}.$$}
 It is easy to verify that the slope of each random variable in
 this linear expression reflects its causal effect, as we obtained in Example \ref{exa:agg1}. We emphasize that for non-linear
aggregate queries \blue{this argument????} would not work, and higher-order correlations should be considered.
\ignore{ It follows that $\mathcal{E}_{t,\mc{Q}}=b-a$. That is,
$\mathcal{E}_{t,\mc{Q}}$ is the slope of of the regression line between the conditional expectation $E(Q |, \nit{E}(X_t))$
and the random variable $X_t$.}
}

\begin{example} \label{exa:eff2} (ex. \ref{exa:agg1} cont.) The  Pearson's correlation coefficients between
the variables $X_{R(n)}$ and the aggregate $\mc{Q}'$ as a random variable are: \ $r^D_{X_{R(450)},\mc{Q}'}=0.9091373,   r^D_{X_{R(150)},   \mc{Q}'}=0.3030458, r^D_{X_{R(100)},  \mc{Q}'}  =0.2020305$, and
$r^D_{X_{R(-100)},\mc{Q}'}=-0.2020305$. \boxtheorem
\end{example}

\vspace{1mm}Causal effect accounts only for the ``linear interaction" between a tuple and  a query answer.  More specifically, it  computes  the
shift of the mean of a query answer due to inserting/deleting a tuple into/from a database (on the space of random interventions on the remaining tuples).  However, inserting/deleting a tuple might change higher-order moments of the query answer distribution.

Causal effect can properly deal with FO queries (due to the multi-linearity of their lineages) and linear aggregate queries. To deal with non-linear aggregate queries, we plan to use information theoretic approaches to quantify causal influence \cite{Janzing14}.

\section{Conclusions and Related Work} \label{sec:degintro}

\ignore{Causal responsibility  as used in \cite{Meliou2010a} intends to provide a
numerical degree of the causal contribution of a
 tuple to a query answer. This responsibility-based ranking is considered as one of the
 most important contributions of HP-causality to data management \cite{Meliou2011a}.} \ignore{Causal responsibility can be traced back to \cite{Chockler04}, where it is
suggested that, for variables $A$ and $B$,  the degree of responsibility of $A$ for $B$ should be  $\frac{1}{(N + 1)}$,  and
$N$ is the {\em minimum number of changes} that have to be made to obtain a situation where $B$
counterfactually depends directly on $A$.}

In \cite{Gerstenberg10} it is argued
that people use something similar to the intuition behind degree of responsibility (in the sense of \cite{Chockler04}) to
 ascribe responsibilities.\ignore{In particular, they ran an experiment and found that
the observed people attributions of responsibility in the experiment were strongly correlated with
the intuition behind degree of responsibility as defined in \cite{Chockler04}.}
In \cite{Zultan12}, it is pointed out that people
 take into account not only the number of changes required to
make $A$ a counterfactual cause for $B$, but also the number of ways to reach a situation where $B$
counterfactually depends on $A$.
In \cite{Halpern15b} it is claimed that, while causal responsibility (in the sense of \cite{Chockler04}) does capture some natural
intuitions,  still alternative definitions might be more appropriate for some applications.

Not surprisingly, much research on causal responsibility can be found in law literature \cite{Wright01,Braham09}\ignore{,Hart59,Moore99,Feinberg68}. However,
in no numerical quantification has been proposed, except for the work  of \cite{Braham09}.\ignore{, which
introduces the notion of ``degree of causality".\ignore{, on the basis of the ``Necessary Element
of a Sufficient Set" (NESS) test.} }

In \cite{Salimithesis}, the notion of degree of causal contribution has been introduced in the context of databases. This notion is defined based on the number of contingency sets associated to a tuple and shown to be closely
related to the proposal in \cite{Braham09} and confirms the intuition behind \cite{Zultan12}. It is not difficult to show that the notion of causal effect as introduced in this paper generalizes that of \cite{Salimithesis}.

\end{document}